# DBNN: Neural Spike Classification Using a Deep Binarized Neural Network

Binyi Ren[1], Luca M. Meyer[1], and Majid Zamani[1]

[1]The School of Electronics and Computer Science (ECS), University of Southampton, Southampton SO17 1BJ, U.K.

br1y24@soton.ac.uk, L.M.Meyer@soton.ac.uk, M.Zamani@soton.ac.uk

**Abstract.** Implantable brain–computer interfaces require on-node spike sorting to reduce telemetry bandwidth and power while maintaining reliable neural decoding. This paper presents a hardware-oriented deep binarized neural network (DBNN) spike-sorting system with two binarized hidden layers with 256 neurons and a fixed-point output layer to enable multiplier-free inference dominated by sign-controlled accumulation and bit-wise logic. The proposed classifier operates on compact 16-sample spike waveforms to reduce the implementation cost (16–256–256–3) and achieves a median classification accuracy of 98.7% on both synthetic and in-vivo datasets. An FPGA prototype on a Cyclone V device operates at 50 MHz and requires 528 cycles per spike, corresponding to a 0.01 ms compute latency, while consuming 828 ALMs and 1023 registers with zero DSP blocks. For ASIC feasibility, the DBNN is implemented using FreePDK45-based flow; synthesis in Synopsys Design Compiler indicates an estimated silicon area of 0.014 mm² and an operating power of 122 nW at 20 kHz under a 1.1 V supply. These results demonstrate that the proposed DBNN spike sorter offers a favorable trade-off between accuracy and implementation cost, supporting low-power, implantable neural interfaces. Overall, the proposed DBNN spike sorter achieves high accuracy (98.7%) with extremely low hardware cost (0.014 mm², 122 nW at 20 kHz) and multiplier-free operation, making it suitable for low-power, implantable neural interfaces. This paper introduces the first DBNN designed for real-time neural spike sorting, striking an excellent balance between input data size and network complexity.

## 1 Introduction

Implantable brain–computer interfaces (iBCIs) increasingly rely on on-chip neural signal processing to translate high-rate extracellular recordings into reliable control variables under strict constraints on power dissipation, silicon area, and end-to-end latency [1, 2]. A key component in this pipeline is spike sorting [3-5], which assigns detected action potentials to their originating neurons and directly impacts both downstream decoding performance and the amount of data that must be communicated off-chip for further processing.

Conventional spike sorting is typically a multi-step pipeline [1, 2] consisting of: (a) pre-processing (band-pass filtering the raw voltage signals); (b) spike detection (often by simple voltage thresholding or more sophisticated non-linear energy operators); (c) spike alignment (e.g. shifting each detected waveform to a common reference point, typically its peak); (d) feature extraction (reducing dimensionality via methods like principal component analysis (PCA) or independent component analysis (ICA)); and (e) clustering or classification of spikes into units (using algorithms such as k-means or hierarchical clustering). Alongside this traditional pipeline, template-matching methods and modern machine learning approaches have been applied [6, 7]. In recent years, deep neural networks (DNNs) have shown excellent accuracy and robustness by learning directly from raw spike waveforms – they can detect spikes, extract features, and perform classification in a unified model. Most neural network spike sorters, however, are not designed with in-situ hardware constraints in mind. Full-precision DNNs, while accurate, require substantial memory for parameters and perform many expensive floating-point multiply–accumulate (MAC) operations [8]. Deploying such models on resource-constrained implantable devices is challenging: it would typically imply large on-chip memory or bandwidth to off-chip RAM, high dynamic power for MAC computations, and difficulties guaranteeing real-time latency across many channels. These issues are exacerbated in wireless iBCIs, where streaming raw data for external processing costs significant power and bandwidth. On-chip spike sorting, therefore, greatly reduces data rate by transmitting only discrete spike trains instead of full waveforms.

Binarized neural networks (BNNs) offer an attractive algorithm–hardware co-design path for implantable and portable systems because binarizing hidden-layer weights and activations enables multiplier-free inference dominated by simple bit-wise operations and greatly reduces parameter storage [9-12]. In 2021, Valencia and Alimohammad introduced a simple binarized neural network (BNN) [13] for classifying neural spikes. Their model, using a spike waveform with 64 samples and a hidden layer of five units, achieved an average accuracy of 90% when tested with synthetic data. In 2023, Valencia and Alimohammad also introduced a new solution using partially binarized neural network (PBNN) [14]. They used DD-Extrema [15] for feature extraction that reduced the network input from 64 to 4 samples. Their model included one hidden layer and an output layer with three units each. They tested different quantization methods and found that the model without binarization had the highest average accuracy of 88.1%, while the binarized model had a lower accuracy of 69.69%.

Building on this principle, this paper proposes hardware implementation of a deep binarized neural network (DBNN) [16] proposed by Meyer et al. in 2025 for supervised spike classification, aiming to retain strong discriminative capability through network depth while preserving the computational regularity and memory efficiency that are critical to low-power integrated implementations [17-19]. However, aggressive quantization can increase training sensitivity and reduce robustness in difficult noise and waveform-overlap regimes; therefore, batch normalization is incorporated during training to stabilize the distribution of pre-activations and improve convergence, while its inference-time effect can be absorbed into thresholding logic to preserve hardware simplicity.

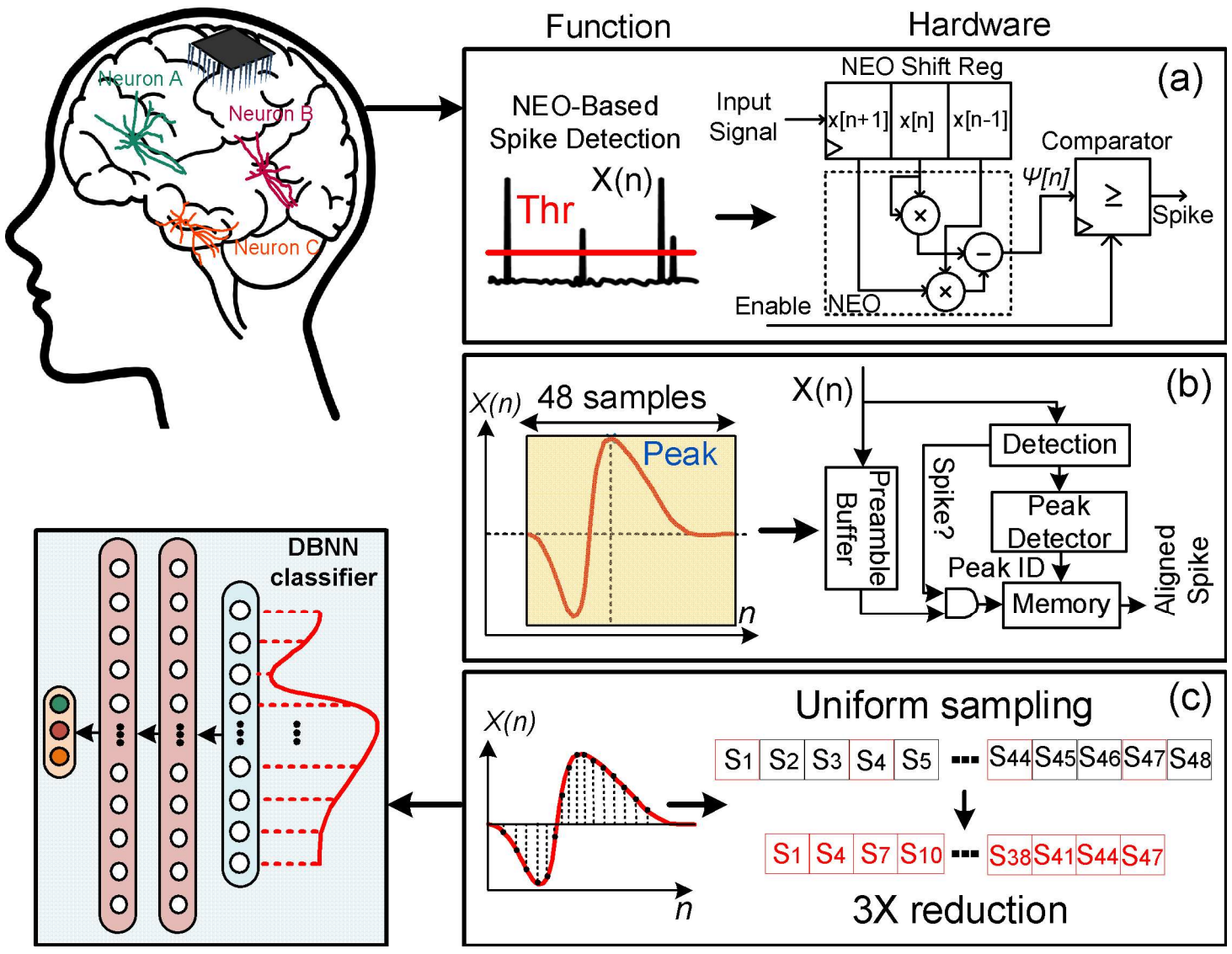

**Fig. 1.** Block diagram of the utilized spike sorting process.

The rest of this paper is structured as follows. Section II discusses the pre-classification pipeline for spike detection, alignment and dimensionality reduction. Section III elaborates on the proposed DBNN architecture and its training procedure for spike classification. Section IV details the hardware implementation of the DBNN classifier core. Section V describes the synthetic and in-vivo datasets used for performance evaluation. Section VI presents the FPGA and ASIC implementation results, alongside a benchmark comparison with prior works. Section VII concludes the paper and discusses future research directions.

## 2 Pre-classification

Fig. 1 shows the entire block diagram of a proposed spike sorting pipeline. The nonlinear energy operator (NEO) [20] in Fig. 1(a) calculates the energy $\Psi(n)$:

$$\Psi(n) = x^2(n) - x(n+1).x(n-1). \qquad (1)$$

for each input sample $x(n)$. During the initial signal acquisition phase, the threshold-calculation module calculates the threshold as:

$$Thr = C\frac{1}{N}\sum_{n=1}^{N}\Psi(n). \qquad (2)$$

where $N$ is the number of samples per window, and $C$ is a constant (empirically chosen to be 8 in this implementation). To reduce the buffering of the threshold calculation, the detection threshold is updated per window rather than per sample.

When a spike is detected ($x(n) > Thr$), 64 samples (18 pre-threshold-crossing or "preamble" samples and 46 post-threshold-crossing samples) of raw data, corresponding to a duration of 2.5 ms, are saved in the preamble buffer shown in Fig. 1(b). For each extracted waveform $x = \{x_i\}_{i=1}^{64}$, the maximum peak $p = \arg\max_{1 \le i \le 64} |x_i|$ is used to align the spikes to a temporal reference. Following alignment, each waveform is cropped to a 48-sample orange window centered on the aligned peak as shown in Fig. 1(b). $p$ serves as the 16th sample of the 48-sample long aligned spike. In the context of spike sorting, dimensionality reduction is the process of reducing the number of features that will be used to cluster the data. This step is essential as it considerably decreases the memory requirements and computational complexity associated with clustering. As shown in Fig. 1(c), each aligned spike waveform of 48 samples is down-sampled to 16 points to 3X reduce dimensionality while preserving shape information. This is essentially the same as choosing evenly spaced samples, $\{S_1, S_2 \ldots S_{48}\}$ to $\{S_1, S_4 \ldots S_{47}\}$.

# 3 Deep Binarized Neural Network for Spike Sorting

In contrast to conventional artificial neural networks (ANNs) that use multi-bit (e.g., floating-point) weights and activations, the deep binarized neural network (DBNN) constrains these parameters to ±1 (bipolar values). By applying the sign function $sign(k) = +1$ if $k \ge 0$ and $-1$ if $k < 0$, the network's synaptic weights and neuron outputs are binarized. This drastic simplification yields significant memory and computation savings: storage requirements shrink nearly by an order of magnitude, and expensive multiply-accumulate operations can be replaced with XNOR and bit-count operations in hardware. For example, storing the weights of a given network in 1-bit form can reduce on-chip memory by 87.5% compared to an 8-bit implementation. Moreover, computation is greatly accelerated since XNOR + popcount (i.e. a function that counts the number of set bits (1s) in the binary representation of an integer) bit-operations are extremely efficient in digital logic. These advantages make BNNs especially attractive for resource-constrained, real-time spike sorting in implantable or portable brain–machine interfaces.

In this paper, a deep fully-connected BNN architecture [16] with an input layer, two hidden layers, and an output layer denoted by 16–256–256–3 for input–hidden1–hidden2–output neurons is adopted for hardware implementation. This topology effectively captures spike waveform features, as increasing network depth/width narrows the accuracy gap between binarized and full-precision models. The input layer has 16 nodes, corresponding to a 16-sample spike after dimensionality reduction. The 16-point vector is normalized to $[-1,1]$ and quantized to 8-bit integers for input to the network. The two hidden layers each comprise 256 binary neurons.

During forward propagation, the weights and activations in the hidden layers are constrained to ±1. Let $W^{(0)}, W^{(1)}$ denote the weight matrices of the first and second hidden layers, respectively, rather than implying only two individual weights. No explicit bias terms are used in the hidden layers, since adding a constant offset is ineffective under the sign activation function. Instead, any bias-like effect is absorbed into the batch normalization (BN) parameters and the corresponding decision thresholds. Each hidden neuron computes a binarized weighted sum, implemented efficiently in hardware using bitwise XNOR operations followed by a population count (popcount), which counts the number of '1' bits in the XNOR result and thereby realises an efficient approximation of the dot product between binarised inputs and weights. BN layers are inserted after each hidden layer during training to stabilize and accelerate convergence. The BN layers adjust and normalize the hidden activations by learning the parameters β and γ, which are used for scaling and centering. This process effectively serves as a bias in a manner that is compatible with binarized units.

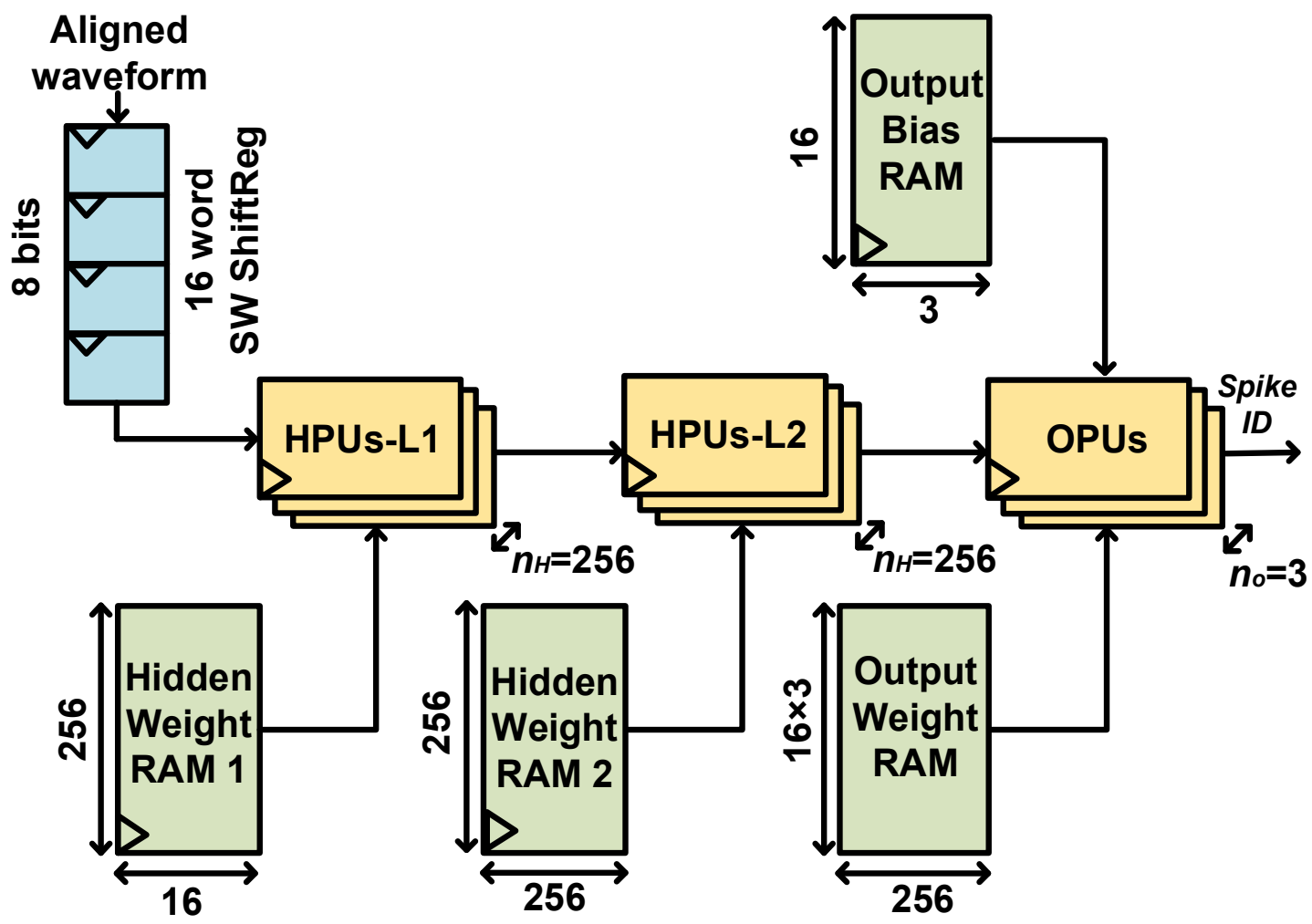

**Fig. 2.** Top-level block diagram of the DBNN-based spike sorting hardware.

At inference time, the learned BN parameters are folded entirely into the decision threshold of the subsequent activation layer, with no modification to the fixed ±1 binarized weights of the preceding layer, incurring only negligible hardware overhead. The final output layer has three neurons since there are three spike classes in the datasets. Unlike the hidden layers, the output layer remains at higher precision: it uses full-precision weights and a bias term to produce three real-valued logit scores. A Softmax activation function is applied to these logits during training to obtain class probabilities, but in deployment the Softmax is omitted. Since the Softmax function is strictly monotonic with respect to its inputs, the argmax of the three raw output logits yields the exact same predicted class as the argmax over the Softmax outputs, without extra computation or a dedicated Softmax hardware unit. Fig. 2 shows the top-level architecture of this DBNN spike sorting classifier. The dimensionality reduced spike waveform is stored in the shift register SW ShiftReg, the depth is set to 16 and each of the registers is 10 bits wide. The HPU-L1 and HPU-L2 represent hidden processing units in layer 1 and 2 respectively. The OPUs also show the output layer processing units. Once the dimensionality reduced spike is received, the control unit begins to read the weight values from weight RAMs (RAM1 and RAM2), output weight and bias weights RAMs to perform the class ID inference.

The classification output of the DBNN can be encoded in different ways. In a one-hot scheme, the network produces one output neuron per class. For three classes, the target outputs are e.g. (1,0,0) for class 1, (0,1,0) for class 2, (0,0,1) for class 3. In a binary encoding scheme, the DBNN uses a minimal number of output bits to represent the classes; specifically, $\lceil log_2 C \rceil$ output neurons can encode $C$ classes. For $C = 3$ classes, this yields 2 output neurons, where, for example, class 3 would be encoded as binary "11" which corresponds to [+1, +1] in the BNN outputs. One-hot encoding is straightforward and was used in the DBNN implementation yielding three one-bit outputs that directly indicate the predicted class. However, the binary encoding is an interesting alternative that reduces the number of output neurons and was reported to provide a form of regularization. In prior work [13], a BNN with binary-coded outputs actually achieved slightly higher sorting accuracy under noise compared to one-hot, since forcing the network to produce distinct bit patterns for each class can help regularize the learning process.

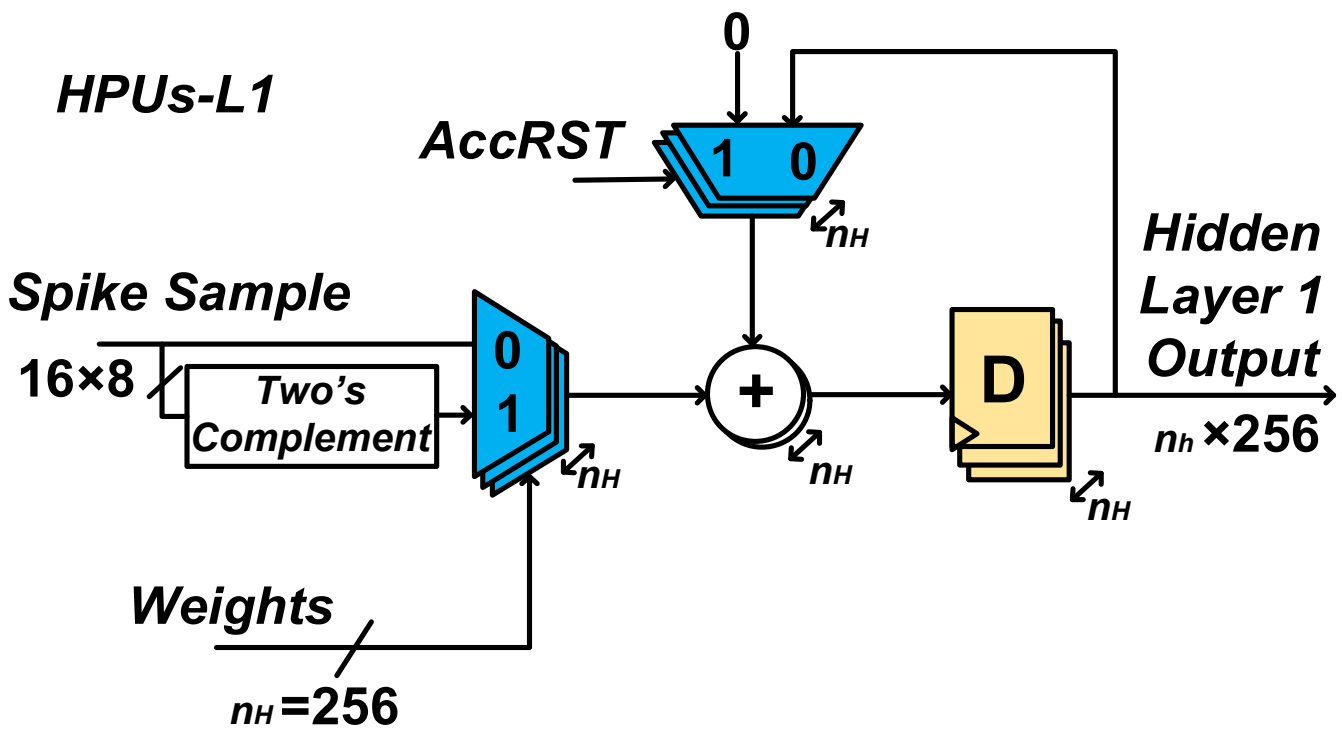

**Fig. 3.** Hidden processing units in Layer 1 (HPUs-L1). The weight values are retrieved from the RAM1 in the HPUs-L1.

The DBNN was trained using supervised learning with labeled spike data. Each input is a vector of 16 features (8-bit integers) representing a single spike waveform, and the ground truth label is one of three neuron classes. The class labels were one-hot encoded. Training is performed using the standard categorical cross-entropy loss. The Adam optimizer was used with an initial learning rate of 0.01 for efficient gradient-based optimization. The network was trained for up to 250 epochs, which was found sufficient for convergence. To prevent overfitting, an early stopping strategy was employed. If the validation loss did not improve for 25 consecutive epochs, training was stopped and the best model was retained. A mini-batch size of 128 spikes was used for each training step, which provided a good balance between gradient stability and memory usage.

## 4 Hardware Architecture and Implementation of the DBNN-Based Classifier

This section details the hardware implementation of the DBNN inference datapath, focusing on the two binarized hidden layers and the fixed-point output layer. All datasets are represented by a 16-sample spike waveform, quantized to signed 8-bit values, and the implemented network topology is therefore fixed to 16–256–256–3. The hardware realization follows the mathematical model in this section, with binarized weights and activations in the hidden layers and higher-precision accumulation in the output layer to preserve classification fidelity.

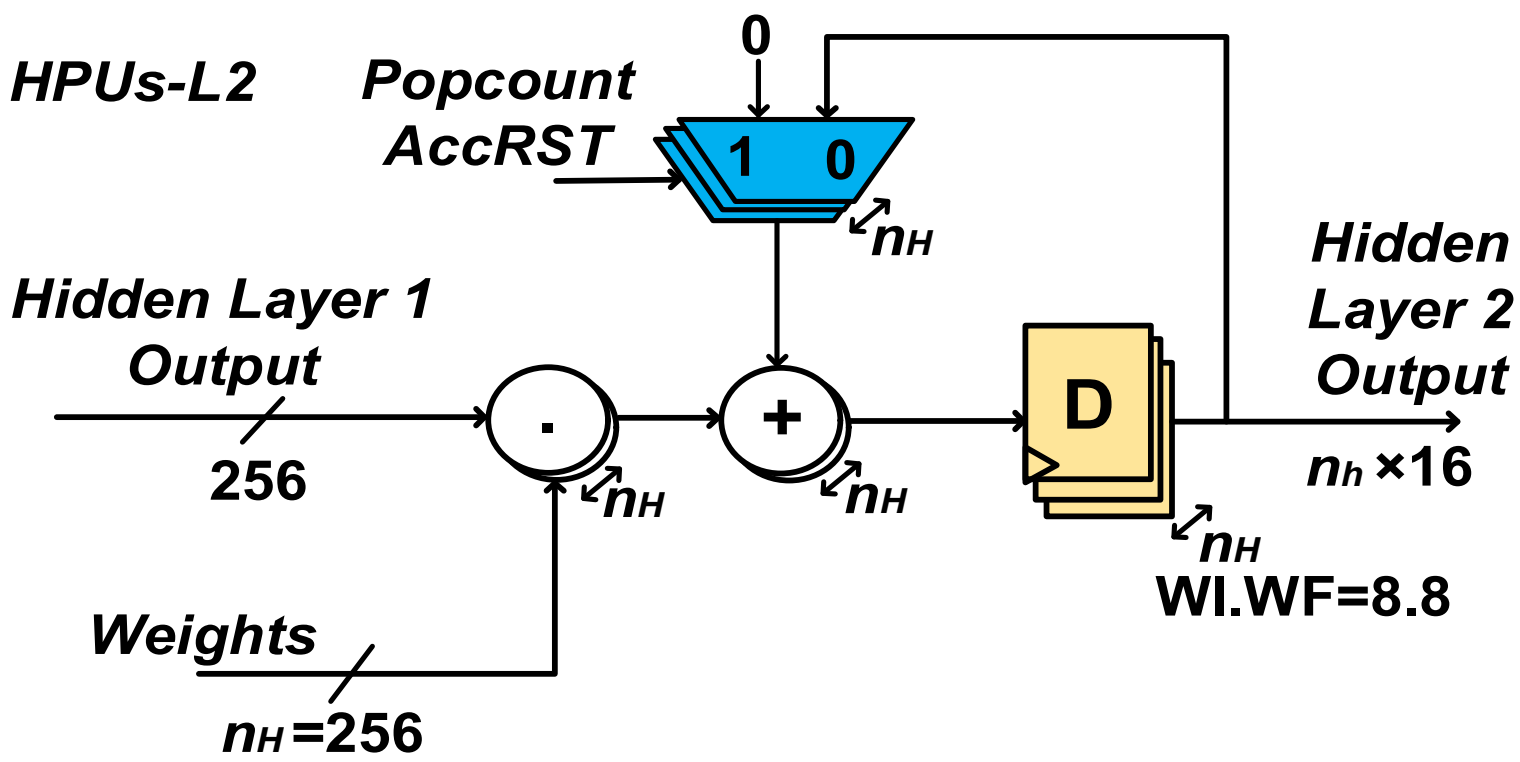

**Fig. 4.** Hidden processing units in Layer 2 (HPUs-L2).

### 4.1 Hidden Processing Units in Layer 1 (HPUs-L1): 8-bit Inputs With 1-bit Weights

The first hidden layer computes $n_H$=256 neuron pre-activations from the 16-point input vector $x_i(i = 1, \ldots, 16)$. Each weight is stored as a single bit $w_{j,i}^{(1)} \in \{0,1\}$ for neuron $j$, corresponding to a bipolar weight $\widetilde{w}_{j,i}^{(1)} \in \{-1, +1\}$ via:

$$\widetilde{w}_{j,i}^{(1)} = 2w_{j,i}^{(1)} - 1. \qquad (3)$$

The pre-activation is:

$$s_j^{(1)} = \sum_{i=1}^{16} \widetilde{w}_{j,i}^{(1)} x_i. \qquad (4)$$

Fig. 3 shows the corresponding datapath. Since $\widetilde{w}_{j,i}^{(1)} \in \{-1, +1\}$, multiplication is implemented as sign selection rather than arithmetic multiplication: a two's-complement unit forms $-x_i$, and a weight-controlled multiplexer selects $+x_i$ or $-x_i$ before accumulation. Consequently, Eq. (2) is implemented using only add/subtract operations and an accumulator, avoiding the use of the multipliers in the first hidden layer. The control signal AccRST is used to reset the accumulator registers. To process each of $nH$ hidden layer neurons in one clock cycle, the HPUs-L1 consists of a time-multiplexed set of multiplexers and accumulators, producing $nH$ bits of output over 256 clock cycles. The hidden activation is obtained by thresholding the accumulated sum $s_j^{(1)}$:

$$a_j^{(1)} = \begin{cases} 1, & s_j^{(1)} \geq T_j^{(1)} \\ 0, & s_j^{(1)} < T_j^{(1)} \end{cases}. \qquad (5)$$

where $T_j^{(1)}$ is a neuron-dependent threshold in layer1. When BN is enabled during training, its affine parameters can be folded into the inference threshold, so the hardware retains the same "accumulate then compare" structure shown in Fig. 3. This is particularly important in binarized networks because BN stabilizes the distribution of pre-activations before the sign function and can substantially improve accuracy in challenging conditions, while still permitting a low-overhead inference implementation through threshold folding.

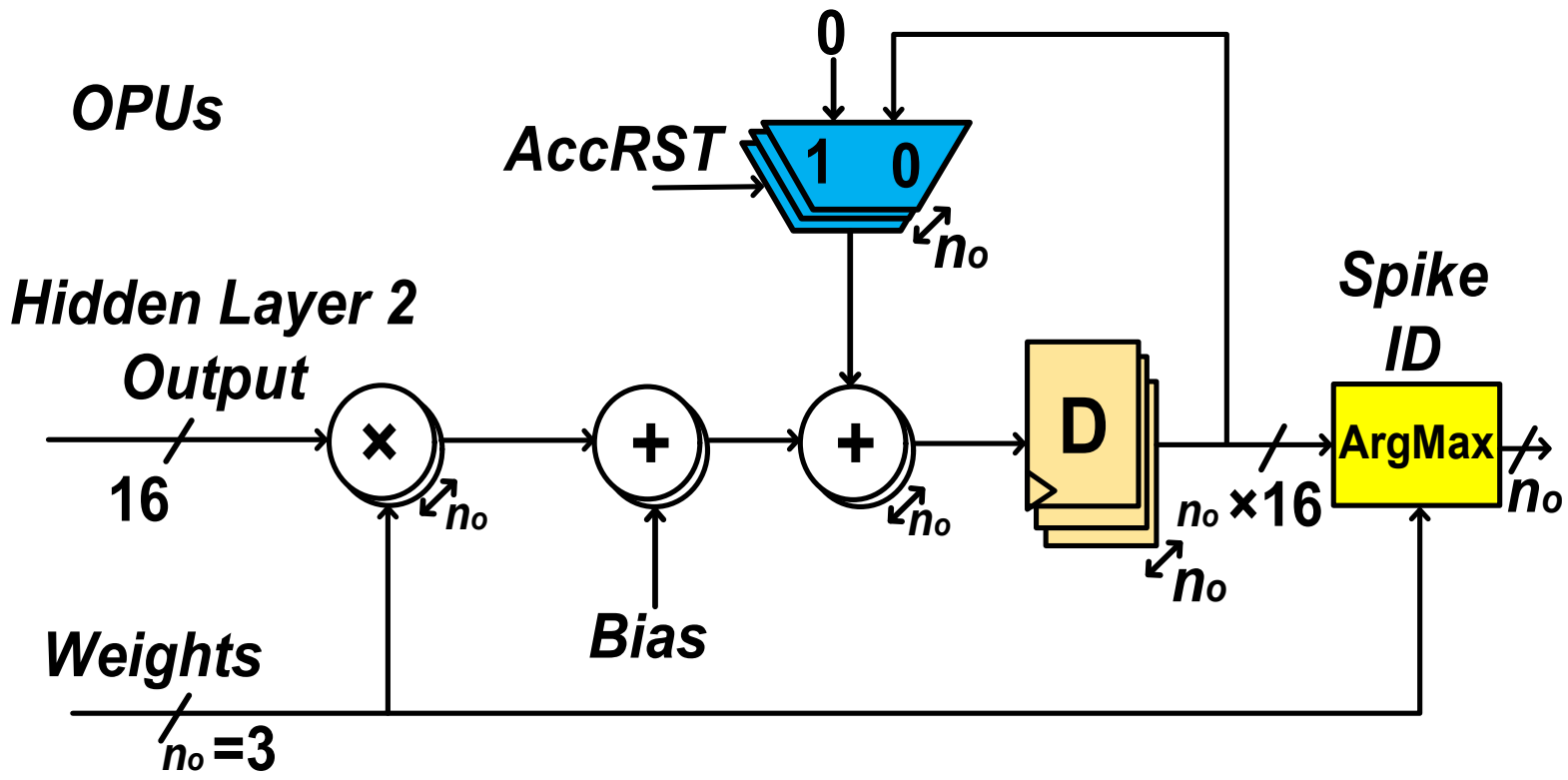

Fig. 5. Output layer processing units (OPUs).

### 4.2 Hidden Processing Units in Layer 2 (HPUs-L2): XNOR–Popcount Binary Dot Product

The second hidden layer receives the 256-bit activation vector $a^{(1)}$ and computes 256 binary outputs. Because both the inputs and the weights are binary, the dot product is expressed using an XNOR–popcount formulation. There are nH neurons in the hidden layer. For neuron k, the number of matching bits is expressed as:

$$p_k = popcount\left(XNOR\left(a^{(1)}, w_k^{(2)}\right)\right), \quad (6)$$

where $w_k^{(2)} \in \{0,1\}^{256}$ denotes the binary weight vector. The popcount accumulates the weighted input feature vectors using 9-bit accumulators and the result is then mapped to a signed correlation:

$$m_k = 2p_k - 256, \quad (7)$$

which is equivalent to a bipolar inner product in which matches contribute $+1$ and mismatches contribute $-1$. Fig. 4 illustrates the correspondent architecture in the second hidden layer. The XNOR stage performs bitwise agreement checking, and the popcount compresses the 256-bit result into an integer sufficient for threshold comparison. The WI.WF denotes the integer and fraction lengths of the accumulation registers, respectively. The activation in layer 2 is generated as:

$$a_k^{(2)} = \begin{cases} 1, & m_k \geq T_k^{(2)} \\ 0, & m_k < T_k^{(2)} \end{cases}, \quad (8)$$

where, analogously to hidden layer 1, BN is added into $T_k^{(2)}$. Compared with HPUs-L1, HPUs-L2 eliminates multi-bit arithmetic almost entirely and relies on logic primitives (XNOR and counting), making it well suited to FPGA look-up table (LUT) fabrics and highly efficient in ASIC implementations.

### 4.3 Output Layer Processing Units (OPUs): Fixed-Point Accumulation and ArgMax Decision

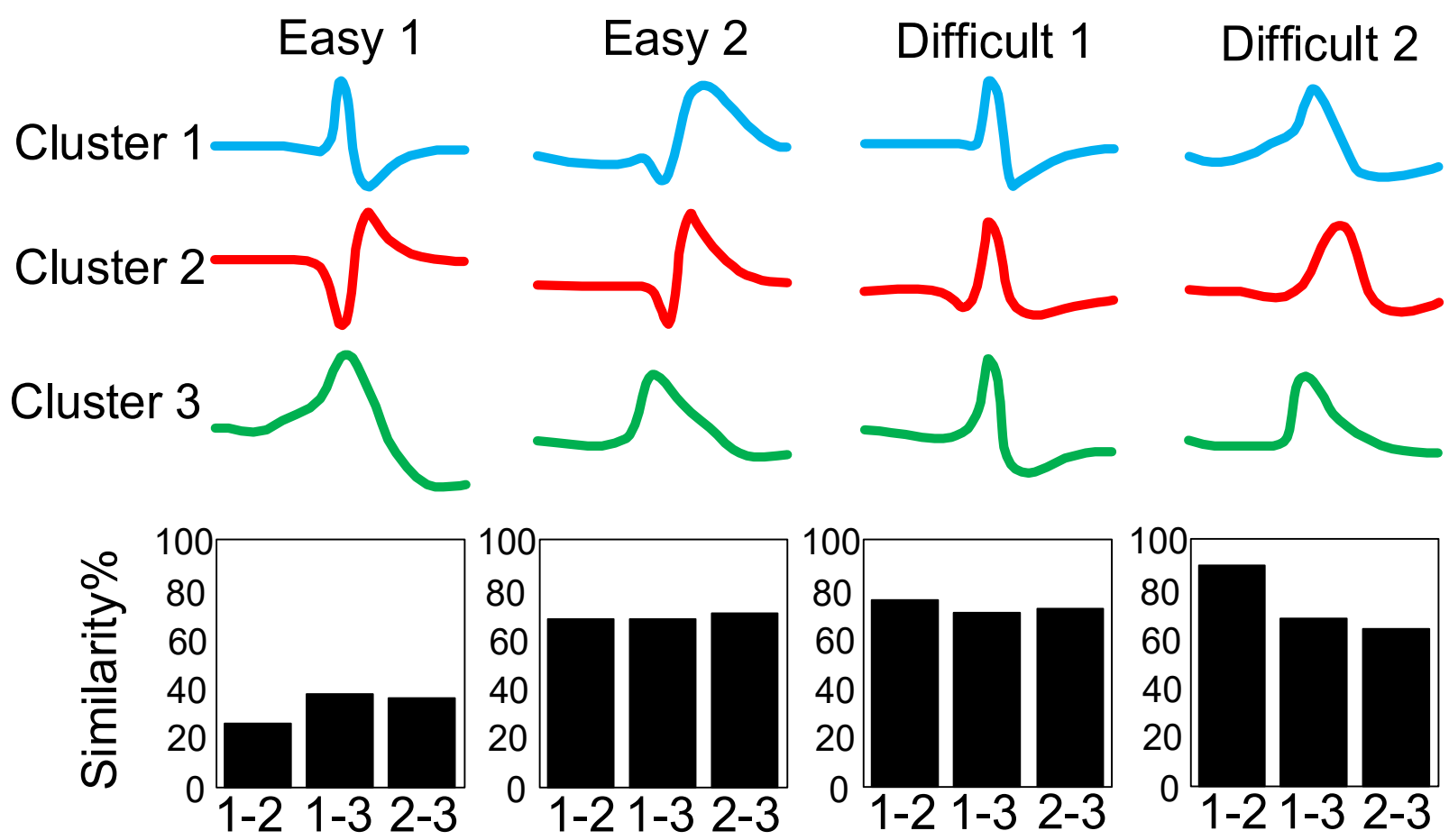

**Fig. 6.** Single unit waveforms used within each dataset. Shown are the mean (peak-aligned) templates with the degree of separability, measured using the Bray–Curtis [21] similarity index (shown at bottom).

The output layer maps the final hidden activations $a^{(2)} \in \{0,1\}^{256}$ to three class scores $\{y_c\}_{c=1}^{3}$. In contrast to the hidden layers, the output weights are stored in fixed-point format to reduce the accuracy loss associated with fully binarized end-to-end inference. Using bipolar activations $\tilde{a}_k^{(2)} = 2a_k^{(2)} - 1$, the class score is computed as:

$$y_c = b_c + \sum_{k=1}^{256} \tilde{a}_k^{(2)} W_{c,k} = b_c + \sum_{k=1}^{256} (2\, a_k^{(2)} - 1) W_{c,k}, \qquad (9)$$

where $W_{c,k}$ and $b_c$ are the fixed-point weight and bias for class $c$. Fig. 5 summarizes the OPUs pipeline. Eq. (9) enables a sign-controlled accumulation implementation: when $a_k^{(2)} = 1$ the datapath adds $W_{c,k}$, and when $a_k^{(2)} = 0$ it subtracts $W_{c,k}$ (implemented via two's complement). After computing the three scores, spike IDs are generated by:

$$\hat{c} = \underset{c \in \{1,2,3\}}{arg\,max}\, y_c, \qquad (10)$$

This is shown by the ArgMax block in Fig. 5. After computing the three scores, spike IDs are generated using the argmax operation defined in Eq. (10).

# 5 Datasets

This section describes the datasets used to evaluate the proposed DBNN spike classifiers and summarizes the corresponding classification results. Two datasets are considered: (1) a synthetic dataset with known ground-truth labels and controlled noise levels, and (2) an in-vivo primate extracellular dataset with manually curated spike labels. In both cases, the preprocessing pipeline is designed to generate compact fixed-length waveform vectors that map directly to the DBNN input layer described in Section II.

**Table 1.** The average classification accuracy (CAcc) of the DBNN over four different datasets and various noise levels without (WOB) and with batch normalization (WB)

| Dataset | | $\sigma_N$ (5%) | $\sigma_N$ (10%) | $\sigma_N$ (15%) | $\sigma_N$ (20%) |
|---|---|---|---|---|---|
| Easy1 | WOB | 99.46% | 99.74% | 99.48% | 99.65% |
| | WB | 99.63% | 99.74% | 99.68% | 99.83% |
| Easy2 | WOB | 99.12% | 99.01% | 97.89% | 96.09% |
| | WB | 99.79% | 99.40% | 98.74% | 98.27% |
| Difficult1 | WOB | 97.00% | 96.34% | 84.97% | 80.96% |
| | WB | 99.62% | 99.41% | 95.34% | 93.07% |
| Difficult2 | WOB | 99.35% | 99.34% | 98.78% | 98.43% |
| | WB | 99.79% | 99.60% | 99.60% | 99.26% |

### 5.1 Synthetic Dataset

The synthetic dataset consists of multiple simulated single-channel recordings, each containing spikes originating from three distinct neurons with full ground-truth spike identities [3]. Many research groups used the available ground-truth labels to train their supervised deep learning models [22-25]. Following the organization used in prior BNN spike-sorting work, the synthetic recordings are grouped into four subsets, Easy1, Easy2, Difficult1, and Difficult2, which differ in waveform separability and the level of spike overlap as shown in Fig. 6. Within each subset, four additive noise conditions are applied, with noise standard deviation ($\sigma_N$) set to 5%, 10%, 15%, and 20% of the spike amplitude.

Spike detection and feature formation follow a consistent procedure as shown in Fig. 1. Each spike event is monitored from the continuous trace using a fixed window of 64 samples. Because overlap is common in the difficult subsets, each 64-sample waveform is aligned to the peak of the dominant spike to eliminate timing offsets from the classification problem. The aligned waveform is then down-sampled by simple decimation to obtain a 16-sample representation that preserves the principal shape characteristics while greatly reducing input dimensionality (i.e. 3X dimensionality reduction). Finally, the 16-sample vector is normalized to the range [−1, 1] and quantized to 8-bit integers, producing a 16-byte input feature vector per spike with an associated class label (neuron 1, 2, or 3).

Tables 1 and 2 report the average classification accuracy (CAcc) of the DBNN on the four synthetic subsets under the four noise conditions without BN and with BN, respectively. CAcc is the number of truly assigned feature vectors (TAFV) over the total number of feature vectors (TNFV), CAcc = (TAFV/ TNFV) × 100%.

First, the DBNN achieves near-ceiling accuracy on the easier subsets across all noise conditions, irrespective of BN. For example, in Easy1 the accuracy remains approximately 99.5–99.8% with or without BN; similarly, Difficult2 also remains close to 99% across noise levels, suggesting that this subset is not strongly degraded by the added noise under the present configuration. Second, the most challenging subset is Difficult1, where accuracy decreases substantially as noise increases when BN is not used. In Table I, Difficult1 drops from 97.00% (5% noise) to 84.96% (20% noise). Considering BN, the same subset achieves markedly higher accuracy at every noise level, improving from 99.62% (5% noise) to 93.07% (20% noise). This demonstrates that BN is particularly beneficial when waveform clusters are poorly separated and the effective signal-to-noise ratio (SNR) is low conditions which is typical in realistic spike sorting.

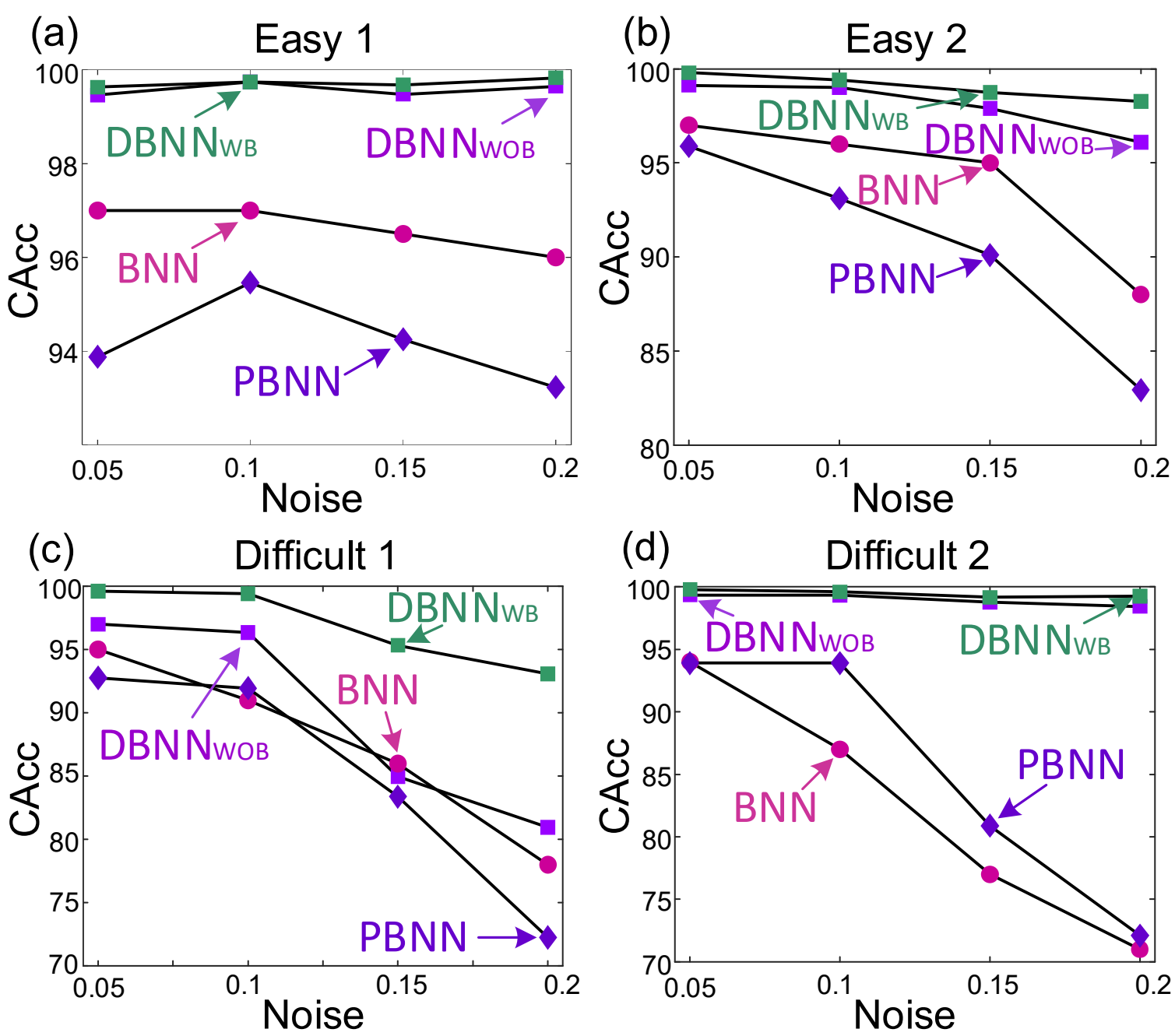

**Fig. 7.** Comparison of the classification accuracy (CAcc) between the proposed DBNN with BNN [13] and PBNN [14]. $DBNN_{WOB}$ and $DBNN_{WB}$ refers to DBNN without and with batch normalization.

The results in Fig. 7 show the comparison of CAcc between DBNN, BNN [13] and PBNN [14]. DBNN with BN (i.e. $DBNN_{WB}$) generally achieves higher CAcc than the other methods. This means that the $DBNN_{WB}$ exhibits the highest immunity to the noise levels and similarity of the spike waveforms amongst all methods.

### 5.2 In-Vivo Dataset

The *in-vivo* dataset consists of multichannel extracellular recordings from primate visual cortex (V1), with spike times and neuron labels obtained via manual sorting [26, 27]. For evaluation, single-channel segments are extracted and processed independently, and a separate DBNN model is trained per channel to account for channel-specific waveform distributions and the number of active neurons.

Due to the structure of the in-vivo recordings and their preprocessing in the pipeline, each spike is represented using 16 samples, which form the DBNN input vector in this case. As with the synthetic dataset, these 16-point vectors are normalized to $[-1, 1]$ and quantized to 8-bit integers, resulting in a 16-byte feature vector per spike. Channels are constrained to a maximum of three labelled neurons; events outside these labels are discarded to preserve a consistent three-class classification formulation. The selected channels represent different difficulty levels in the spike waveforms.

Table 2 summarizes the classification accuracy on eight representative channels and reports accuracy both with BN and without BN, together with the absolute improvement in the last column. The results show that BN yields consistent and substantial gains for most channels. In particular, channels that are challenging without BN (e.g., channel 85 at 72.36% and channel 69 at 76.62%) improve to 99.46% and 98.21% with BN, respectively. In contrast, channels that are already well-separated including channels 76 and 80 show only marginal improvement because baseline accuracy is already near saturation. On average, BN improves the in-vivo classification accuracy from a channel-averaged 84.16% to 98.23% across the 8 evaluated channels, corresponding to a mean absolute improvement of 18.04%.

**Table 2.** The classification accuracy of the DBNN over in-vivo datasets

| Channel | $DBNN_{WOB}$ | $DBNN_{WB}$ | Improvement |
|---|---|---|---|
| 66 | 78.71% | 97.05% | 23.30% |
| 69 | 76.62% | 98.21% | 28.18% |
| 76 | 98.97% | 99.93% | 0.97% |
| 77 | 77.18% | 96.93% | 25.59% |
| 79 | 84.33% | 96.27% | 14.16% |
| 80 | 98.21% | 99.35% | 1.16% |
| 84 | 86.90% | 98.61% | 13.48% |
| 85 | 72.36% | 99.46% | 37.45% |
| Average | 84.16% | 98.23% | 18.04% |

# 6 Implementation Results

This section presents the outcomes of implementing the proposed DBNN-based spike-sorting classifier on FPGA and ASIC in Synopsys-Cadence, and compares the design with notable spike-sorting hardware documented in existing literature.

The DBNN classifier is motivated by the necessity for hardware-friendly spike classification in implantable neural interfaces: deep-learning spike sorting is dominated by dot products which can be power consuming, while binarized inference largely reduces computation to bitwise operations and accumulation, enabling more resource-efficient on-chip processing.

## 6.1 FPGA implementation results

The DBNN classifier was prototyped on an Intel Cyclone V FPGA, with performance evaluated both for the core inference engine and the full test setup including a UART host interface. For the hardware core alone, spike classification completes in 528 clock cycles at 50 MHz, yielding a compute latency of 10.56 $\mu s$ (0.01 ms) and a maximum throughput of ~100 kspikes/s. When including UART communication overhead (115.2 kbps baud rate for transmitting 16-byte input waveforms and 1-byte class labels), the end-to-end test latency increases to ~1.5 ms per spike. The UART-induced delay is purely a test interface artifact and would not be present in implantable deployments, where spike data is fed directly from on-chip analog front-end (AFE) circuits with negligible latency.

From a real-time deployment perspective, the latency from receiving the final input sample to producing the output class is approximately 0.01 ms, which is negligible compared with typical closed-loop BCI timing requirements (<1 ms). These latency and resource utilization measurements are consistent with the architectural design intent detailed in Section IV: all core arithmetic operations in the hidden layers are reduced to sign-controlled accumulation (first layer) and XNOR–popcount (second layer), eliminating the need for any DSP blocks.

Resource utilization after FPGA implementation indicates that the complete spike sorter including UART and control consumes 828 adaptive logic modules (ALMs) out of approximately 32k available on the Cyclone V, and 1023 registers, which is a small fraction of the target device capacity. This low utilization is consistent with the DBNN design principle emphasized namely, replacing multiply–accumulate (MAC) operations in the hidden layers with sign-controlled accumulation and bit-wise primitives (e.g., XNOR/popcount or add/subtract selection), thereby avoiding expensive arithmetic blocks.

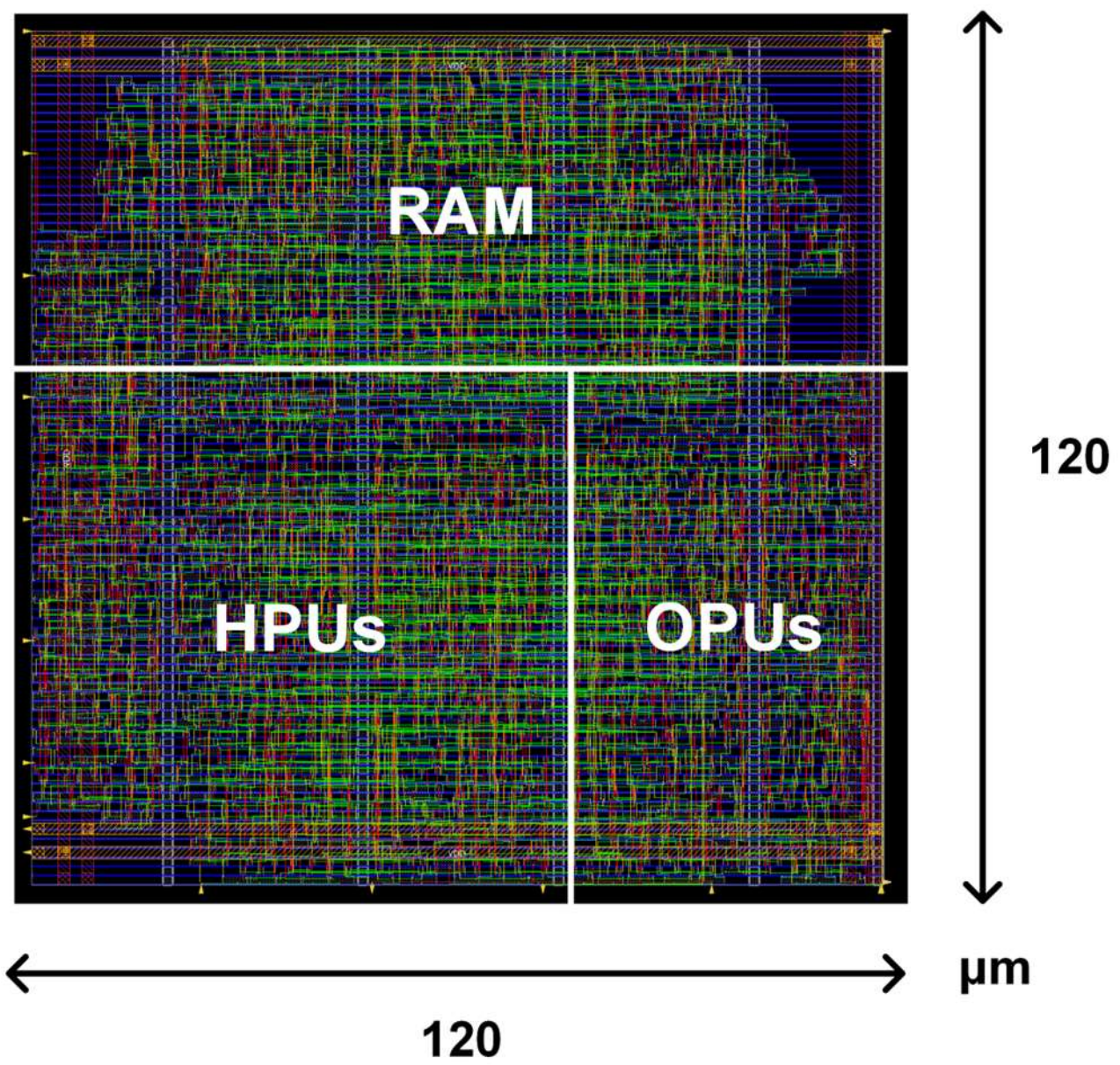

**Fig. 8.** The layout of the designed DBNN-based spike classifier.

### 6.2 ASIC synthesis results

The proposed DBNN-based spike-sorting system has been implemented using the FreePDK45 process design kit to assess ASIC feasibility. As illustrated in Fig. 8, the chip layout of the DBNN spike-sorting core is expected to occupy approximately $0.014\, mm^2$ of silicon area. Logic synthesis was performed using Synopsys Design Compiler, while place-and-route and sign-off timing analysis are being conducted in Cadence Innovus (post-layout results to be reported when available). At an operating frequency of 20 kHz and a supply voltage of 1.1 V, the estimated power consumption of the BNN-based spike-sorting ASIC is $0.122\, \mu W$ . The FPGA and ASIC implementations employ an identical micro-architectural datapath and control strategy; the principal distinction lies in the underlying hardware realization. Specifically, the FPGA implementation maps shift-register structures onto vendor-specific LUT-based resources, whereas the ASIC implementation realizes the same sequential elements using positive-edge-triggered standard-cell flip-flops.

To assess the feasibility of the proposed spike-sorting system for brain-implantable operation, key telemetry-related metrics including the wireless transmission power and the resulting power density of ASIC were computed. During spike sorting, the sampling rate is $r_s = 20$ kS/s. With a 16-bit representation of the neural signal, the corresponding input data rate is $r_{\text{in}} = 320$ kb/s. Assuming a typical neuronal firing rate of 40 spikes/s and encoding the spike-train output using 3 bits per event, the output data rate becomes $r_o = 120$ b/s. Hence, the proposed DBNN-based spike-sorting system achieves a data-rate reduction of approximately $\frac{r_{\text{in}}}{r_o} \approx 2667\, \times$.

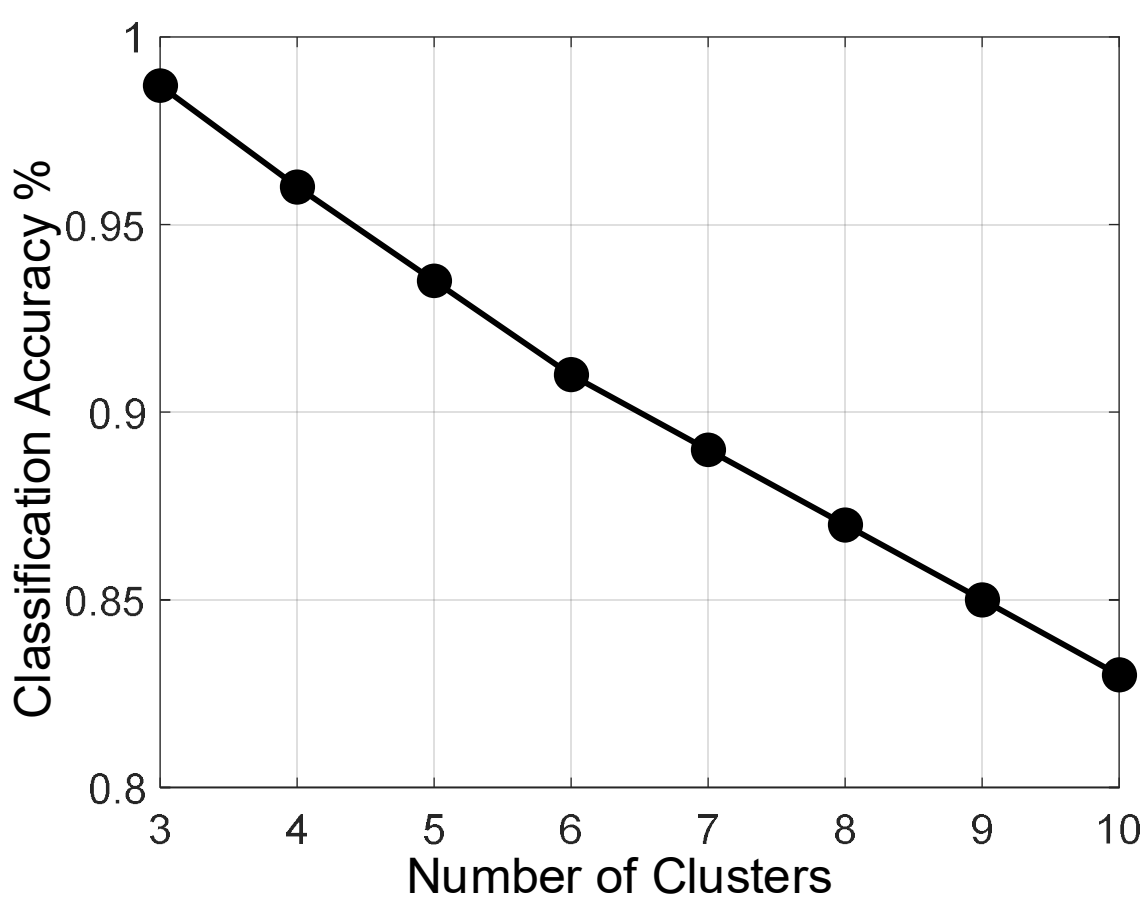

**Fig. 9.** Classification accuracy of the $DBNN_{WB}$ utilizing varying numbers of clusters from 3 to 10 waveforms.

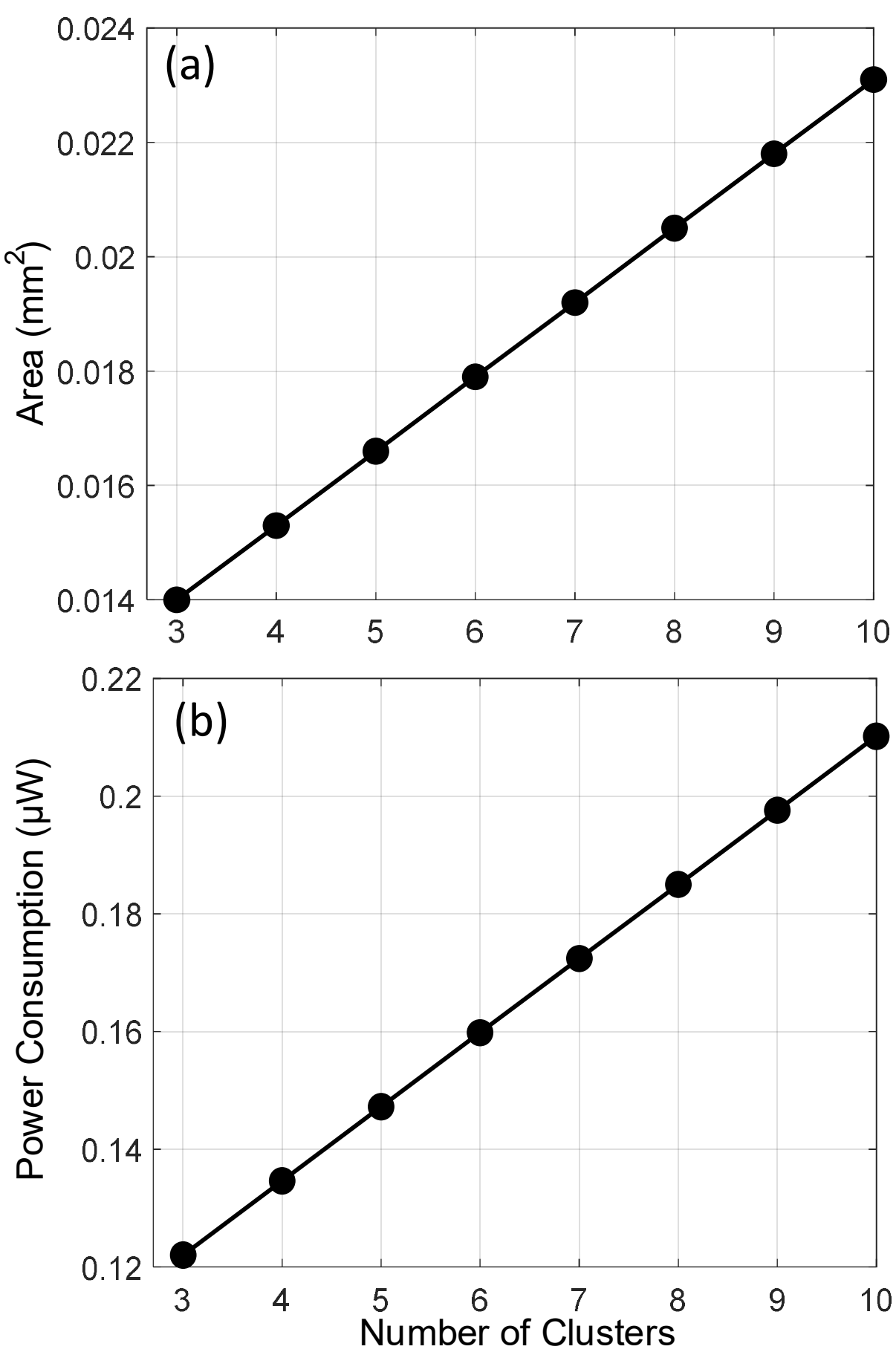

**Fig. 10.** (a) Area utilization and (b) power consumption of the ASIC with storage for 3 to 10 template waveforms.

This substantial reduction in output throughput directly lowers the required wireless transmission power, thereby reducing the energy budget associated with telemetering the sorted spike trains.

It is important to highlight that the above data-rate reduction is enabled by the on-chip supervised classification of the proposed DBNN, which maps raw spike waveforms to discrete neuron identity labels, rather than transmitting full continuous neural recordings, while maintaining state-of-the-art classification accuracy. Based on the energy-per-bit reported in (approximately 3 nJ/bit [28]), the transmission power at $r_o$ is estimated as $P_{\text{tx}} \approx 3$ nJ/bit $\times$ 120 b/s $\approx 0.36\ \mu W$. Accordingly, the total ASIC power is $0.482\mu$W. For a core area of $0.014\ mm^2$, this corresponds to a power density of $\frac{34.4\ \mu W}{mm^2}$ , which remains within commonly accepted tissue-safety limits for implantable devices.

To explore the impact of the number of classes on the implementation results of the ASIC, Fig. 9 shows CAcc of the DBNN utilizing varying numbers of clusters from 3 to 10 clusters. Fig. 10(a) and Fig. 10(b) also show the area utilization and power consumption for supporting 3 to 10 neuron classes. The results show that classification accuracy decreases roughly linearly with the increasing number of classes, while area utilization and power consumption increase approximately linearly with the number of supported classes.

### 6.3 Comparative analysis (ASIC and FPGA)

To contextualize the proposed DBNN implementation, two comparison tables are presented that contrast the design with ASIC and FPGA implementations.

A range of ASIC spike-sorting implementations has been reported in the literature. Table 3 summarizes the key characteristics and implementation outcomes of representative state-of-the-art systems. As indicated in Table 3, the proposed DBNN-based spike sorting system requires a drastically reduced silicon area and consumes less power than compared designs, while delivering outstanding classification accuracy that outperforms most prior binarized neural network implementations and conventional spike-sorting approaches. In [29], a parallel OSort-based spike-sorting ASIC was designed and implemented in a 32-nm CMOS process. In [6], a template- matching (TM) spike-sorting ASIC employing three templates was realized in a 45-nm CMOS technology. The design in [30] performs NEO-based spike detection, aligns detected spikes to the maximum derivative, and extracts features via discrete derivatives; it integrates four 16-channel modules that output aligned waveforms or feature vectors but does not carry out real-time classification. In [31], spikes are detected using an absolute-value criterion and clustered using online sorting (OSort) within a single 16-channel module. Similarly, [32] applies voltage-threshold detection followed by alignment to the maximum absolute amplitude and OSort-based clustering. The work in [20] reports single-channel spike sorting with NEO-based detection, maximum-amplitude alignment, and derivative-based feature extraction, supporting an unsupervised learning process. Also, the work in [33] presents a multi-channel spike-sorting ASIC architecture centred on discrete wavelet transform (DWT) feature extraction with decision tree classifier. Finally, the proposed work in [34] presents the ASIC implementation of a dictionary learning-based feature extraction. It uses dictionary values limited to the values set of [−1, 0, +1], multiplication-free informative feature extraction. The work in [34] implements Hadamard matrix, equiangular tight

**Table 3.** The ASIC characteristics and implementation results of various spike sorting systems

| Design | Algorithm | Supervised | Technolog y(nm) | VDD (V) | Operating frequency (kHz) | Area ($mm^2$) | Power per channel ($\mu W$) | Accuracy | Data rate reduction |
|---|---|---|---|---|---|---|---|---|---|
| Ours | DBNN | Y | 45 | 1.1 | 20 | 0.014 | 0.122[a, b] | 0.987 | 2667× |
| [13] | BNN | Y | 180 | 1.8 | 24 | 0.33 | 2.02 | 0.91 | 2000× |
| [14] | PBNN | Y | 180 | 1.8 | 24 | 0.34 | 2.5 | 0.91 | 2000× |
| [29] | OSort | N | 32 | 1.16 | 24 | 2.57 | 2.78 | 0.87 | 1600× |
| [6] | TM | Y | 45 | 0.25 | 24 | 0.30 | 0.064 | 0.90 | 3200× |
| [30] | FE[d] | Y | 90 | 0.55 | 4000 | 0.06 | 2.03 | 0.77 | 11× |
| [31] | OSort | N | 45 | – | 56 | 0.077 | 10.3[c] | 0.93 | 278× |
| [32] | OSort | N | 65 | 0.27 | 480 | 0.07 | 4.68 | 0.75 | 240× |
| [20] | CSort | Y | 180 | 1.8 | 960 | 2.7 | 148 | 0.85 | 150×/240× |
| [33] | FE | Y | 130 | 1.2 | 160 | 0.023 | 0.75 | 0.85 | – |
| [34] | Dictionary | N | 65 | 1 | 30 | 0.09 | 0.181[c] | 0.92 | – |

a- The power consumption of NEO, peak alignment and uniform sampling is 0.09 μW.
b- Using lower supply voltage in a more advanced technology results in lowest dynamic power consumption. For example, scaling voltage from 1.1V to 0.45V introduces $(1.1/0.45)^2 = 5.97$ reduction in dynamic power (20.5 nW).
c- Note that the authors only list the ASIC implementation results for the feature extraction module.
d- [30] and [32] report silicon-verified measurement results.

frame (ETF) and random Bernoulli matrix low-cost dictionaries. As indicated in Table 3, the DBNN-based spike sorting system necessitates a reduced silicon area and utilizes less power which is compatible with implantable devices constraints, while delivering outstanding classification accuracy.
The efficiency advantages of the proposed DBNN become more evident when comparing hardware metrics. Table 4 shows the features and outcomes of different spike sorting systems implemented on FPGAs. The DBNN FPGA implementation used <3% of logic resources and no DSP blocks for a single-channel sorter and scaled linearly for more channels. In contrast, a full-precision DNN or even a 4-bit quantized network would consume far more multipliers and memory. For example, an equally-sized full-precision network (256-256 hidden units) would require on the order of 65k 16-bit multiplications per spike – which on an FPGA would either use dozens of DSP blocks time-multiplexed at a high clock or results in a much longer latency. By binarizing, multiplications are converted into simple XNOR operations and addition, allowing the design to run at a comfortable 50 MHz with ample headroom.
[13] proposed the BNN-based spike classifier, using a single 5-neuron hidden layer with binarized weights. They achieved ~90% accuracy on synthetic data but struggled on more difficult waveforms. The proposed DBNN, with two 256-neuron hidden layers, substantially outperforms that earlier BNN – thanks to its deeper architecture and using BN, it handles noise and overlapping spikes far better (achieving ~98.7% on similar data vs. ~90%). In [35], an OSort-based spike-sorting system was designed on Zynq-7000. Differently, [36] used an L-Sort system on Zynq MPSoC. [6] presented TM algorithm on Virtex-6, whose latency was only 0.55 $\mu s$. Similarly, authors in [37] proposed the hardware implementation of a bayes-optimal TM-based spike sorting system. While the maximum operating frequency of their design was not reported, their sorting latency was given as 53 clock cycles. In [38], a probabilistic neural network (PNN) based spike-sorting algorithm was implemented on a Virtex-6 FPGA, achieving a reported classification accuracy of 92%. However, because the evaluation was conducted on a non-public dataset, a direct quantitative comparison with our results is not possible. In [39], the authors proposed a real-time spike classification system based on Hebbian learning, implemented on a Spartan-6 FPGA. In that design, principal component analysis (PCA) is employed to project the input spike waveforms onto a reduced feature space prior to classification. Experimental results were reported for both 10-bit and 16-bit word lengths, and the fixed-point behavior of the system was modelled using the MATLAB Fixed-Point Toolbox.

# 7 Conclusion

The proposed DBNN-based spike-sorting classifier achieves a median classification accuracy of 98.7% on both synthetic and in-vivo datasets while consuming only 122 nW and 0.014 mm² in a 45 nm technology. By binarizing the hidden layers, the proposed approach realizes a hardware-oriented inference data path dominated by bit-wise operations and sign-controlled accumulation, while preserving a higher-precision output stage to maintain decision fidelity. The study evaluates the classifier on both synthetic data with controlled conditions and in-vivo recordings to demonstrate practical relevance, and it combines FPGA prototyping with ASIC-oriented synthesis analysis to support feasibility for deployment under tight power/area/latency constraints. Future work will explore extending the DBNN to support real-time adaptation to signal instabilities, such as electrode drift or increases in noise levels, and for high-density microelectrode array processing.

**Table 4.** The FPGA characteristics and implementation results of various spike sorting systems

| Design | Algorithm | FPGA device | Clock / Max Freq. (MHz) | Latency | Logic | Registers | BRAM | DSP | Accuracy |
|---|---|---|---|---|---|---|---|---|---|
| Ours | DBNN | Cyclone V | 50 | 10.56 $\mu s$ | 828[a] | 1023 | 11 | 0 | 98.7% |
| [13] | BNN | Artix-7 | 125 | 0.51 $\mu s$ | 314 | 267 | 2 | 0 | 93% |
| [35] | O-Sort | Zynq-7000 | 200 | 7.2 ms | 51674 | 17484 | 98 | 60 | - |
| [36] | L-Sort | Zynq MPSoC | 3.6 | 33 $\mu s$ | 6513 | 3017 | 6 | 15 | 96% |
| [6] | TM | Virtex-6 | 122 | 0.55 $\mu s$ | 6635 | 4880 | 0 | 5 | 90% |
| [37] | TM | Virtex-6 | - | 2.65 $\mu s$ | 190000 | 29000 | 24 | - | 96% |
| [38] | PNN | Virtex-6 | 100 | 6.7 $\mu s$ | 17712 | 3936 | 7 | 175 | - |
| [39] | Hebbian | Spartan-6 | 78 | 0.96 $\mu s$ | ~6216 | ~4662 | 41 | - | 95% |

a- The total design with UART uses 827.5 ALMs and 1023 registers; DBNN core only uses 725 ALMs and 853 registers.

**Data Availability**: Data sharing is not applicable to this article due to ongoing research and development involving the deep binarized algorithm outlined in this paper.

## References

[1] Meyer, L., Zamani, M., Rokai, J., Demosthenous, A.: Deep learning based spike sorting: a survey. J. Neural Eng. (2024). https://doi.org/10.1088/1741-2552/ad8b6c

[2] Zhang, T., Rahimi Azghadi, M., Lammie, C., Amirsoleimani, A., Genov, R.: Spike sorting algorithms and their efficient hardware implementation: a comprehensive survey. J. Neural Eng. 20(2), 021001 (2023). https://doi.org/10.1088/1741-2552/acc7cc

[3] Quiroga, R.Q., Nadasdy, Z., Ben-Shaul, Y.: Unsupervised spike detection and sorting with wavelets and superparamagnetic clustering. Neural Comput. 16(8), 1661–1687 (2004). https://doi.org/10.1162/089976604774201631

[4] Okreghe, C.O., Zamani, M., Demosthenous, A.: A deep neural network-based spike sorting with improved channel selection and artefact removal. IEEE Access 11, 15131–15143 (2023)

[5] Fang, T., Zamani, M.: A meta-fusion architecture for few-shot classification of spike waveforms in high bandwidth brain-machine interfacing. arXiv preprint (2025). https://doi.org/10.48550/arXiv.2503.18040

[6] Valencia, D., Alimohammad, A.: An efficient hardware architecture for template matching-based spike sorting. IEEE Trans. Biomed. Circuits Syst. 13(3), 481–492 (2019). https://doi.org/10.1109/TBCAS.2019.2907882

[7] Meyer, L.M., Zamani, M.: MetaSort: An accelerated approach for non-uniform compression and few-shot classification of neural spike waveforms. arXiv preprint (2026)

[8] Sze, V., Chen, Y.-H., Yang, T.-J., Emer, J.S.: Efficient processing of deep neural networks: a tutorial and survey. Proc. IEEE 105(12), 2295–2329 (2017). https://doi.org/10.1109/JPROC.2017.2761740

[9] Hubara, I., Soudry, D., Yaniv, R.E.: Binarized neural networks. arXiv preprint (2016). https://doi.org/10.48550/arXiv.1602.02505

[10] Courbariaux, M., Bengio, Y., David, J.-P.: BinaryConnect: training deep neural networks with binary weights during propagations. arXiv preprint (2015). https://doi.org/10.48550/arXiv.1511.00363

[11] Rastegari, M., Ordonez, V., Redmon, J., Farhadi, A.: XNOR-Net: ImageNet classification using binary convolutional neural networks. In: Comput. Vis. – ECCV 2016, LNCS, vol. 9908, pp. 525–542. Springer, Cham (2016). https://doi.org/10.1007/978-3-319-46493-0_32

[12] Simons, T., Lee, D.-J.: A review of binarized neural networks. Electronics 8(6), 661 (2019). https://doi.org/10.3390/electronics8060661

[13] Valencia, D., Alimohammad, A.: Neural spike sorting using binarized neural networks. IEEE Trans. Neural Syst. Rehabil. Eng. 29, 206–214 (2021). https://doi.org/10.1109/TNSRE.2020.3043403

[14] Valencia, D., Alimohammad, A.: Partially binarized neural networks for efficient spike sorting. Biomed. Eng. Lett. 13(1), 73–83 (2023). https://doi.org/10.1007/s13534-022-00255-7

[15] Zamani, M., Demosthenous, A.: Feature extraction using extrema sampling of discrete derivatives for spike sorting in implantable upper-limb neural prostheses. IEEE Trans. Neural Syst. Rehabil. Eng. 22(4), 716–726 (2014)

[16] Meyer, L.M., Zamani, M., Demosthenous, A.: Binarized neural networks for resource-efficient spike sorting. IEEE Access 13, 60258–60269 (2025). https://doi.org/10.1109/ACCESS.2025.3552008

[17] Umuroglu, Y. et al.: FINN: A framework for fast, scalable binarized neural network inference. In: Proc. ACM/SIGDA FPGA, pp. 65–74 (2017). https://doi.org/10.1145/3020078.3021744

[18] Blott, M. et al.: FINN-R: An end-to-end deep-learning framework for fast exploration of quantized neural networks. ACM Trans. Reconfigurable Technol. Syst. 11(3), 1–23 (2018). https://doi.org/10.1145/3242897

[19] Conti, F., Schiavone, P.D., Benini, L.: XNOR neural engine: a hardware accelerator IP for binary neural network inference. IEEE Trans. Comput.-Aided Des. Integr. Circuits Syst. 37(11), 2940–2951 (2018). https://doi.org/10.1109/TCAD.2018.2857019

[20] Zamani, M., Jiang, D., Demosthenous, A.: An adaptive neural spike processor with embedded active learning for improved unsupervised sorting accuracy. IEEE Trans. Biomed. Circuits Syst. 12(3), 665–676 (2018). https://doi.org/10.1109/TBCAS.2018.2825421

[21] Lian, J., Garner, G., Muessig, D., Lang, V.: A simple method to quantify the morphological similarity between signals. J. Signal Process. 90(2), 684–688 (2010)

[22] Li, Z., Wang, Y., Zhang, N., Li, X.: An accurate and robust method for spike sorting based on convolutional neural networks. Brain Sci. 10(11), 835 (2020). https://doi.org/10.3390/brainsci10110835

[23] Liu, M., Feng, J., Wang, Y., Li, Z.: Classification of overlapping spikes using convolutional neural networks and long short-term memory. Comput. Biol. Med. 148, 105888 (2022). https://doi.org/10.1016/j.compbiomed.2022.105888

[24] Li, S., Tang, Z., Yang, L., Li, M., Shang, Z.: Application of deep reinforcement learning for spike sorting under multi-class imbalance. Comput. Biol. Med. 164, 107253 (2023). https://doi.org/10.1016/j.compbiomed.2023.107253

[25] Wang, M. et al.: A deep learning network based on CNN and sliding window LSTM for spike sorting. Comput. Biol. Med. 159, 106879 (2023). https://doi.org/10.1016/j.compbiomed.2023.106879

[26] Chu, C., Chien, P., Hung, C.: Multi-electrode recordings of ongoing activity and responses to parametric stimuli in macaque V1. CRCNS.org (2014). https://doi.org/10.6080/K0J1012K

[27] Chu, C.C.J., Chien, P.F., Hung, C.P.: Tuning dissimilarity explains short distance decline of spontaneous spike correlation in macaque V1. Vision Res. 96, 113–132 (2014). https://doi.org/10.1016/j.visres.2014.01.008

[28] Chen, F., Chandrakasan, A.P., Stojanovic, V.M.: Design and analysis of a hardware-efficient compressed sensing architecture for data compression in wireless sensors. IEEE J. Solid-State Circuits 47(3), 744–756 (2012). https://doi.org/10.1109/JSSC.2011.2179451

[29] Valencia, D., Alimohammad, A.: A real-time spike sorting system using parallel OSort clustering. IEEE Trans. Biomed. Circuits Syst. 13(6), 1700–1713 (2019). https://doi.org/10.1109/TBCAS.2019.2947618

[30] Karkare, V., Gibson, S., Marković, D.: A 130-μW, 64-channel neural spike-sorting DSP chip. IEEE J. Solid-State Circuits 46(5), 1214–1222 (2011). https://doi.org/10.1109/JSSC.2011.2116410

[31] Liu, Y., Sheng, J., Herbordt, M.C.: A hardware design for in-brain neural spike sorting. In: Proc. IEEE HPEC, pp. 1–6 (2016). https://doi.org/10.1109/HPEC.2016.7761590

[32] Karkare, V., Gibson, S., Marković, D.: A 75-μW, 16-channel neural spike-sorting processor with unsupervised clustering. IEEE J. Solid-State Circuits 48(9), 2230–2238 (2013)

[33] Yang, Y., Boling, S., Mason, A.J.: A hardware-efficient scalable spike sorting neural signal processor module for implantable high-channel-count brain-machine interfaces. IEEE Trans. Biomed. Circuits Syst. 11(4), 743–754 (2017). https://doi.org/10.1109/TBCAS.2017.2679032

[34] Zamani, M., Sokolic, J., Jiang, D., Renna, F., Rodrigues, M., Demosthenous, A.: Accurate, very low computational complexity spike sorting using unsupervised matched subspace learning. IEEE Trans. Biomed. Circuits Syst. 14(2), 221–231 (2020)

[35] Schaffer, L., Nagy, Z., Kincses, Z., Fiath, R., Ulbert, I.: Spatial information based OSort for real-time spike sorting using FPGA. IEEE Trans. Biomed. Eng. 68(1), 99–108 (2021). https://doi.org/10.1109/TBME.2020.2996281

[36] Han, Y., Wang, S., Hamilton, A.: L-Sort: An efficient hardware for real-time multi-channel spike sorting with localization. In: Proc. IEEE BioCAS, pp. 1–5 (2024). https://doi.org/10.1109/BioCAS61083.2024.10798317

[37] Dragas, J., Jackel, D., Hierlemann, A., Franke, F.: Complexity optimization and high-throughput low-latency hardware implementation of a multi-electrode spike-sorting algorithm. IEEE Trans. Neural Syst. Rehabil. Eng. 23(2), 149–158 (2015). https://doi.org/10.1109/TNSRE.2014.2370510

[38] Wang, D. et al.: FPGA implementation of hardware processing modules as coprocessors in brain-machine interfaces. In: Proc. IEEE EMBC, pp. 4613–4616 (2011). https://doi.org/10.1109/IEMBS.2011.6091142

[39] Yu, B. et al.: Real-time FPGA-based multichannel spike sorting using Hebbian eigenfilters. IEEE J. Emerg. Sel. Top. Circuits Syst. 1(4), 502–515 (2011). https://doi.org/10.1109/JETCAS.2012.2183430